\documentclass[
    aps,
    prd,
    superscriptaddress,
    nofootinbib,
    amsmath,
    amssymb
]{revtex4-2}

\usepackage[T1]{fontenc}
\usepackage[utf8]{inputenc}
\usepackage{amsmath}
\usepackage{amssymb}
\usepackage{bm}
\usepackage{mathtools}
\usepackage{graphicx}
\usepackage{dcolumn}
\usepackage{booktabs}
\usepackage[
    colorlinks=true,
    linkcolor=blue,
    citecolor=blue,
    urlcolor=blue
]{hyperref}
\usepackage{microtype}

\newcommand{\dd}{\mathrm{d}}
\newcommand{\Cm}{C_{\bar m m\bar m m}}
\renewcommand{\Re}{\operatorname{Re}}
\renewcommand{\Im}{\operatorname{Im}}
\newcommand{\PTT}{P_{\mathrm{TT}}}

\numberwithin{equation}{section}

\begin{document}

\title{Helicity-squared scaling of post-eikonal phases for electromagnetic
and gravitational waves in Ricci-flat spacetimes}

\author{C. A. S. Almeida}
\email{carlos@fisica.ufc.br}
\affiliation{Departamento de F\'{i}sica, Universidade Federal do Cear\'{a},
60455-760 Fortaleza, Cear\'{a}, Brazil}
\date{\today}

\begin{abstract}
Spin-optical descriptions of electromagnetic and gravitational waves
isolate finite-frequency corrections that are odd in the helicity. We
study the complementary, helicity-even part of the first post-eikonal
correction for scalar, electromagnetic, and linearized gravitational
waves on Ricci-flat backgrounds. In a parallel-propagated null tetrad,
the direct Weyl coupling of the gravitational field projects onto the
circular transverse-traceless states as the multiple
$2C_{\bar m m\bar m m}$ of the identity. The Ricci identity for the spin
coefficient $\mu$ then rewrites this projection as a total derivative
plus optical terms. The helicity-even phase of a spin-$s$ wave, measured
relative to the scalar wave, then obeys a helicity-squared law,
$\Delta_2=4\Delta_1$, up to endpoint terms. The Weyl coupling is what completes the $h^2$ scaling in the gravitational case. For
weak-deflection scattering by a Schwarzschild black hole we evaluate the
relevant integrals along the ray and obtain
$\Delta_s=-3\pi s^2/(8M\omega\rho^3)$, with $\rho=b/M$. The same
coefficient follows independently, and without reference to any frame
or gauge, from the Regge--Wheeler phase shifts combined with the
asymptotics of the rotation functions $d^l_{ss}$. Direct numerical
integration of the radial equations confirms it. At first order in $M$
the helicity-even phase difference vanishes, as required by the Born
amplitudes. The helicity-odd spin-optical sector and the helicity-even
post-eikonal sector are therefore organized by the first and second
powers of the helicity, respectively.
\end{abstract}

\maketitle

\section{Introduction}
\label{sec:introduction}

The high-frequency propagation of waves in curved spacetime has been
studied since the geometrical-optics expansion of Isaacson, and the
subject now includes systematic higher-order treatments of scalar,
electromagnetic, and gravitational waves
\cite{Isaacson1968I,Isaacson1968II,Dolan2018,Dolan2018HO,Harte2019}.
Spin optics and Berry geometry gave finite-frequency corrections a
particularly clear form \cite{Berry1984}. For circularly polarized
electromagnetic and gravitational waves, polarization transport
generates helicity-dependent corrections to the ray dynamics, among
them the gravitational spin Hall effect
\cite{FrolovShoom2011,Yamamoto2018,Frolov2020,Oancea2020,Andersson2021,
Dahal2022,HarteOancea2022,AnderssonOancea2023,FrolovShoom2024}. These
methods have been applied to Schwarzschild and Kerr, and the graviton
Berry curvature has recently been obtained from quantum field theory
\cite{FrolovSchwarzschild2024,FrolovKoek2025,Ito2026}.

The leading spin-optical correction is odd under helicity reversal,
$h\to-h$. It is not, however, the whole first post-eikonal correction.
At the same order in the inverse frequency the transport equation for
the amplitude contains three kinds of terms: the Berry term, linear in
the polarization connection and odd in $h$; terms quadratic in the
transverse derivatives of the polarization frame, which are even in $h$;
and, for gravitational waves only, a direct algebraic coupling to the
background Weyl tensor,
\begin{equation}
(\mathcal W b)_{\mu\nu}=2C_{\mu\alpha\nu\beta}b^{\alpha\beta}.
\end{equation}
On a Ricci-flat background Maxwell's equations have no term of this
kind. Spin-optical treatments retain the first group of terms, and the
helicity-even terms are usually left aside, either because they do not
affect the ray trajectory at this order or because, taken separately,
they depend on the choice of frame. Earlier work on polarized waves has
accordingly focused on helicity-odd effects or on the general structure
of the corrected field
\cite{Dolan2018HO,Harte2019,Dahal2023,CusinLagos2020}, and
beyond-eikonal corrections in Schwarzschild have been computed for
scalar waves \cite{BruyerePitrou2026}.

Two questions are left open. The first is how the helicity-even phase
depends on the spin of the field. The second is whether the Weyl
coupling gives gravitational waves a structure with no electromagnetic
analogue at this order. Both bear on a structural statement that is
well established for the helicity-odd sector. There the effective ray
equations of gravitational waves have the same form as those derived
from Maxwell's equations \cite{Andersson2021}, and fields of different
spin differ only by a constant factor, the helicity, in the spin tensor
\cite{AnderssonOancea2023}; methods developed for electromagnetic waves
are carried over to gravitational waves by accounting for the
difference in helicity \cite{Dahal2022}. Whether a comparable statement
holds for the helicity-even sector, where the Weyl coupling is the only
algebraic contact between the spin-2 field and the background
curvature, has not been examined. To our knowledge, the spin dependence
of the helicity-even post-eikonal phase has not been determined.

In this paper we show that the helicity-even sector obeys a
helicity-squared law, and that the Weyl coupling is precisely what makes
this law hold for gravity. Projected onto the circular
transverse-traceless (TT) states, the Weyl operator is $2\Cm$ times the
identity, so it neither converts nor splits the two gravitational
helicities. In a parallel-propagated null tetrad, the Ricci identity for
the spin coefficient $\mu$ rewrites $\Cm$ as a total derivative plus
products of optical scalars. The frame-derivative terms contain a
connection part weighted by $s^2$ and an optical part weighted by $s$.
The two weights coincide for $s=1$, and for $s=2$ the Weyl coupling
supplies exactly their difference. The helicity-even phase of a spin-$s$
wave, measured relative to the scalar wave on the same congruence, is
then $s^2$ times a common bulk integral plus endpoint terms. An analysis
of the field strengths at the detector shows that the endpoint terms
cancel, apart from local terms, so the law holds for the
gauge-invariant phases. The helicity-odd sector is linear in $h$, and
the helicity-even sector is quadratic. In practical terms, the
helicity-even phase of a gravitational wave follows from the
electromagnetic one through the factor $h^2=4$, without a separate
evaluation of the Weyl term.

We then evaluate the helicity-even sector for weak-deflection scattering
by a Schwarzschild black hole, in two independent ways. One calculation
integrates the bulk quantities along the scattered ray. The other uses
the Regge--Wheeler phase shifts and the asymptotics of the rotation
functions $d^l_{ss}$, and involves no choice of frame or gauge. Both give
\begin{equation}
\Delta_s=-\frac{3\pi}{8}\,\frac{s^2}{M\omega\,\rho^3}
+\mathcal O\!\left(\frac{1}{M\omega\,\rho^4}\right),
\qquad \rho=\frac bM ,
\label{eq:main_result}
\end{equation}
where $b$ is the impact parameter and $\omega$ the frequency. The first
order in $M$ cancels, as required by the exact Born amplitudes
\cite{DeLogiKovacs1977}, and direct numerical integration of the radial
equations confirms the second-order coefficient.

The main contributions of this work are therefore the following:
(i) the identity that relates the algebraic Weyl projection to the
optical scalars of the congruence, and the helicity-squared law that
follows from it; (ii) the endpoint analysis that extends the law to the
gauge-invariant field strengths; and (iii) an explicit coefficient,
obtained by two independent methods, which also serves as an analytic
benchmark for numerical treatments of wave propagation in black-hole
spacetimes. The effect itself is far below present observational
sensitivity (Sec.~\ref{sec:discussion}); the interest of the result lies
in the structure it reveals.

The paper is organized as follows.
Section~\ref{sec:equations} collects the post-eikonal equations for
$s=0,1,2$ and the algebraic Weyl projection.
Section~\ref{sec:projection} projects the first post-eikonal source in
a parallel null tetrad. Section~\ref{sec:law} derives the helicity-squared
law and treats the endpoint terms. Section~\ref{sec:schwarzschild_ray}
evaluates the bulk integrals in Schwarzschild, and
Sec.~\ref{sec:partial_waves} gives the partial-wave calculation.
Section~\ref{sec:discussion} relates the result to spin optics and to
wave-optics studies of lensing by rotating black holes, and
Sec.~\ref{sec:conclusions} summarizes. Two appendixes contain the
weak-deflection evaluation of the Weyl integral and the angular phase
of $d^l_{ss}$.

We use signature $(-,+,+,+)$ and the curvature convention
$[\nabla_c,\nabla_b]V^a=R^a{}_{dcb}V^d$, for which the linearized
Einstein equation takes the form \eqref{eq:grav_wave}.

\section{Post-eikonal equations for $s=0,1,2$}
\label{sec:equations}

Consider a Ricci-flat background, $R_{\mu\nu}=0$. A massless scalar obeys
$\Box\phi=0$. In Lorenz gauge the Maxwell potential obeys
$\Box A_\mu-R_\mu{}^\nu A_\nu=0$, which reduces to $\Box A_\mu=0$. For
a traceless Lorenz-gauge metric perturbation, linearized gravity
gives
\begin{equation}
\Box h_{\mu\nu}+2C_{\mu\alpha\nu\beta}h^{\alpha\beta}=0
\label{eq:grav_wave}
\end{equation}
\cite{Isaacson1968I,Dolan2018,Harte2019}. The three cases may be written
together as $\Box u+\mathcal W_s u=0$, with $\mathcal W_0=\mathcal W_1=0$ and
$\mathcal W_2=\mathcal W$.

We use the WKB ansatz
\begin{equation}
u=e^{iS/\epsilon}\left(b_0+\epsilon b_1+\mathcal O(\epsilon^2)\right),
\qquad k_\mu=\nabla_\mu S .
\label{eq:WKB}
\end{equation}
At leading order $k^\mu k_\mu=0$, so the rays are affinely parametrized
null geodesics, $k^\nu\nabla_\nu k^\mu=0$. With the transport operator
\begin{equation}
\mathcal D\equiv 2k^\rho\nabla_\rho+\nabla_\rho k^\rho ,
\end{equation}
the next two orders give
\begin{equation}
\mathcal D b_0=0,
\qquad
i\mathcal D b_1+\Box b_0+\mathcal W_s b_0=0 .
\label{eq:post_eikonal}
\end{equation}
The gauge conditions must be imposed order by order. At leading order they
make $b_0$ transverse (and traceless for $s=2$). At the next order they
fix the longitudinal part of $b_1$, for example
$ik^\mu b_{1\mu\nu}=-\nabla^\mu b_{0\mu\nu}$ for gravity.

Let $m^\mu$ be a complex transverse null vector with $m\cdot\bar m=1$,
parallel transported along the rays. The circular states of helicity
$h=\pm s$ are
\begin{equation}
e^{(+s)}=m^{\otimes s},\qquad e^{(-s)}=\bar m^{\otimes s},
\label{eq:helicity_basis}
\end{equation}
and the leading amplitude is $b_0=\mathcal B\,e^{(h)}$. Here $\mathcal B$
is real and obeys the flux law $2k\cdot\nabla\mathcal B+(\nabla\cdot
k)\mathcal B=0$.

\subsection{Algebraic Weyl projection}

For $s=2$, the matrix elements of $\mathcal W$ between the circular TT
states are
\begin{equation}
\bar e_{(\pm)}^{\mu\nu}\mathcal W_{\mu\nu}{}^{\alpha\beta}e^{(\mp)}_{\alpha\beta}=0,
\qquad
\bar e_{(+)}^{\mu\nu}\mathcal W_{\mu\nu}{}^{\alpha\beta}e^{(+)}_{\alpha\beta}
=\bar e_{(-)}^{\mu\nu}\mathcal W_{\mu\nu}{}^{\alpha\beta}e^{(-)}_{\alpha\beta}
=2\Cm .
\end{equation}
The off-diagonal elements vanish because the Weyl tensor is
antisymmetric in each index pair. The diagonal elements coincide by pair
symmetry, and $\Cm$ is real. Therefore
\begin{equation}
\PTT\,\mathcal W\,\PTT=2\Cm\,\mathbf 1_{\rm TT}.
\label{eq:Weyl_TT}
\end{equation}
The direct Weyl coupling produces neither $+2\leftrightarrow-2$ conversion
nor a splitting of the two helicities: it is helicity even.

The scalar $\Cm$ is invariant under $m\to e^{i\psi}m$. It is not
invariant under a null rotation $m\to m+\zeta k$, which preserves all
inner products of $m$ but mixes in components of the Weyl tensor along
$k$. Such a null rotation corresponds, at leading order, to a residual
gauge transformation of $b_0$. For this reason the integral of $\Cm$
along a ray is not by itself an observable. The next sections show how it
combines with the derivative terms.

\section{Projected transport in a parallel null tetrad}
\label{sec:projection}

\subsection{Frame gradients}

Complete $(k,m,\bar m)$ to a null tetrad $(k,n,m,\bar m)$, with
$k\cdot n=-1$, parallel transported along the rays. Since $k=\nabla S$,
the tensor $\nabla_a k_b$ is symmetric. We use the projections
\begin{equation}
\rho=m^a\bar m^b\nabla_bk_a,\quad
\sigma=m^am^b\nabla_bk_a,\quad
\mu=n_am^b\nabla_b\bar m^a,\quad
\lambda=n_a\bar m^b\nabla_b\bar m^a,
\label{eq:spin_coefficients}
\end{equation}
which correspond to the Newman--Penrose spin coefficients
\cite{NewmanPenrose1962}, up to signature-dependent signs. Here $\rho$ is
real. The polarization connection is
\begin{equation}
A_a=-i\,\bar m^b\nabla_am_b ,
\label{eq:connection}
\end{equation}
a real covector with $k^aA_a=0$ for a parallel frame. Using $m\cdot m=0$,
$m\cdot\bar m=1$ and $k\cdot m=0$, the gradient of $m$ decomposes as
\begin{equation}
\nabla_\rho m^\mu=iA_\rho m^\mu-(n\cdot\nabla_\rho m)\,k^\mu
+(m^\nu\nabla_\rho k_\nu)\,n^\mu .
\end{equation}
Writing $A_m=A_am^a$, it follows that
\begin{equation}
A_aA^a=2|A_m|^2,
\qquad
K_m\equiv\nabla_\rho\bar m_\mu\nabla^\rho m^\mu
=2|A_m|^2+2\Re(\rho\mu+\sigma\lambda).
\label{eq:K}
\end{equation}

\subsection{Projected source}

Projecting $\Box(\mathcal B\,m^{\otimes s})$ onto $\bar m^{\otimes s}$ and
using $\bar m\cdot\Box m=i\nabla\cdot A-K_m$, we obtain
\begin{align}
\bar m^{\otimes s}\cdot\Box(\mathcal B\,m^{\otimes s})&=E_s+iG_s,
\\
E_s&=\Box\mathcal B-\mathcal B\left[sK_m+s(s-1)A_aA^a\right],
\label{eq:E_s}
\\
G_s&=2sA^a\nabla_a\mathcal B+s\mathcal B\nabla_aA^a .
\label{eq:G_s}
\end{align}
For helicity $-s$ the sign of $G_s$ is reversed. The projections onto the
opposite helicity vanish, because $m\cdot m=0$ and
$m\cdot\nabla_\rho m=0$. For $s=2$ these expressions reproduce the
decomposition $D_{\pm\pm}=E\pm iG$, $D_{+-}=D_{-+}=0$ of the
gravitational TT source.

\subsection{Phase and amplitude}

Write the TT component of the first correction as
$\bar m^{\otimes s}\cdot b_1=i\mathcal B\varphi_s$. Since $k\cdot\nabla\bar
m=0$, the projection commutes with $\mathcal D$, and
Eq.~\eqref{eq:post_eikonal} gives
\begin{equation}
\frac{\dd\varphi_s}{\dd\lambda}
=\frac{E_s\pm iG_s}{2\mathcal B}+c_s\,\Cm ,
\qquad c_0=c_1=0,\quad c_2=1 .
\label{eq:phi_rate}
\end{equation}
Since $u\simeq\mathcal B\,e^{iS/\epsilon}(1+i\epsilon\varphi_s)e^{(h)}$, the
real part of $\varphi_s$ corrects the phase and the imaginary part the
amplitude. The amplitude correction $\mp G_s/(2\mathcal B)$ is linear in
$A$ and odd in the helicity. It is the field-level counterpart of the
Berry term responsible for the spin-optical corrections
\cite{Oancea2020,Andersson2021,FrolovShoom2024}. The phase correction is
helicity even.

The scalar term $\Box\mathcal B/(2\mathcal B)$ is common to all spins. It
also dominates the approach to the caustic of the scattered congruence.
We therefore consider the difference $\Delta_s\equiv\Phi_s-\Phi_0$ between
the post-eikonal phases of the spin-$s$ and scalar waves on the same
congruence. From Eqs.~\eqref{eq:K}--\eqref{eq:phi_rate},
\begin{equation}
\frac{\dd\Delta_s}{\dd\lambda}
=\epsilon\left[-s^2|A_m|^2-s\,\Re(\rho\mu+\sigma\lambda)+c_s\Cm\right].
\label{eq:Delta_rate}
\end{equation}

\section{Helicity-squared law}
\label{sec:law}

\subsection{Ricci identity}

Differentiate $\mu$ along the ray and commute the derivatives. Using
$k\cdot\nabla n=k\cdot\nabla m=0$ and
$m^b\nabla_bk^c=\rho\,m^c+\sigma\,\bar m^c-(n\cdot V)k^c$, with
$V^c=m^b\nabla_bk^c$, one finds
\begin{equation}
\frac{\dd\mu}{\dd\lambda}=-(\rho\mu+\sigma\lambda)+C_{n\bar mkm}.
\label{eq:dmu}
\end{equation}
Tracelessness of the Weyl tensor, contracted with
$g^{ab}=-k^an^b-n^ak^b+m^a\bar m^b+\bar m^am^b$, gives
$\Cm=-2\Re\,C_{n\bar mkm}$. Hence
\begin{equation}
\Cm=-2\,\frac{\dd\,\Re\mu}{\dd\lambda}-2\Re(\rho\mu+\sigma\lambda).
\label{eq:Ricci}
\end{equation}
The algebraic projection of the Weyl tensor that enters the gravitational
equation is thus a total derivative plus the same optical combination
that appears in $K_m$.

Two consequences follow at once. First, at order $M$ the products
$\rho\mu$ and $\sigma\lambda$ are of second order. The integral of $\Cm$
along the ray then reduces to the boundary term $-2[\Re\mu]$, which
vanishes for scattering because $\Re\mu$ decays at both ends. This is the
structural reason why the nominal $\rho^{-2}$ contribution of the Weyl
term integrates to zero (Appendix~\ref{app:weyl_integral}).

Second, inserting Eq.~\eqref{eq:Ricci} into Eq.~\eqref{eq:Delta_rate} and
integrating along the ray, we obtain
\begin{equation}
\Delta_s=\epsilon\left\{s^2\left(\tfrac12I_C-I_A\right)
+s\,[\Re\mu]_{\rm in}^{\rm out}
+\left(c_s-\tfrac{s(s-1)}{2}\right)I_C\right\},
\label{eq:Delta_integrated}
\end{equation}
where
\begin{equation}
I_C=\int\Cm\,\dd\lambda,
\qquad
I_A=\int|A_m|^2\,\dd\lambda .
\label{eq:IC_IA}
\end{equation}
The last term vanishes both for Maxwell ($c_1=0$) and for linearized
gravity ($c_2=1$). Therefore
\begin{equation}
\Delta_s=\epsilon\,s^2\left(\tfrac12I_C-I_A\right)+\epsilon\,s\,[\Re\mu]_{\rm in}^{\rm out},
\qquad s=1,2,
\label{eq:h2_law}
\end{equation}
and, in particular,
\begin{equation}
\Delta_2=4\Delta_1-2\epsilon\,[\Re\mu]_{\rm in}^{\rm out}.
\label{eq:Delta2_Delta1}
\end{equation}

Equation~\eqref{eq:h2_law} is the central structural result of this
work. Without the Weyl coupling, the optical term in
Eq.~\eqref{eq:Delta_rate} would carry the coefficient $s$, while the
connection term carries $s^2$. For $s=1$ the two coincide. For $s=2$ the
direct Weyl coupling contributes exactly the missing $s(s-1)/2$ times
$I_C$. The Weyl term is therefore not an independent helicity-even
structure superimposed on the electromagnetic one: it is the term that
makes the helicity-even sector of the gravitational wave scale as $h^2$.
The operator structure is familiar from
$(\nabla-ihA)^2=\nabla^2-2ihA\cdot\nabla-h^2A^2$. The linear term
underlies spin optics, and the quadratic term is its helicity-even
partner.

\subsection{Frame dependence}

The separate integrals in Eq.~\eqref{eq:h2_law} are defined in the
parallel frame fixed on the incident screen. Under a local rotation
$m\to e^{i\psi}m$, with $\mathcal B\to\mathcal B\,e^{-is\psi}$, the field
$b_0$ is unchanged, and so is the total phase. Under a parallel null
rotation $m\to m+\zeta k$ with constant $\zeta$, $I_C$ changes only at
relative order $\rho^{-1}$ (Appendix~\ref{app:weyl_integral}). The same
holds for $I_A$: since $\delta A_m=-i(\zeta\rho-\bar\zeta\sigma)$, with
$\zeta=\mathcal O(M/b)$ and $\rho,\sigma=\mathcal O(M)$, the change is of
the same relative order.

The TT phase $\Phi_s$ is not itself a gauge-invariant quantity. The
invariant observables are field strengths evaluated at the detector,
and they differ from the TT phase by endpoint terms that do not
accumulate along the ray. For Maxwell, the Newman--Penrose scalar
$F_{n\bar m}$ follows directly from Eq.~\eqref{eq:WKB}:
\begin{equation}
F_{n\bar m}=-\frac{i}{\epsilon}\,\mathcal B\,e^{iS/\epsilon}
\left[1+i\epsilon\left(\varphi_1+n\cdot\nabla\ln\mathcal B-\Re\mu\right)
-\epsilon\left(A_n+\Im\mu\right)+\mathcal O(\epsilon^2)\right].
\label{eq:F_nm}
\end{equation}

\subsection{Endpoint terms for the gravitational amplitude}
\label{subsec:grav_endpoint}

Equation~\eqref{eq:F_nm} shows that, for Maxwell, the phase of a
field-strength component is the TT phase plus local terms, and that the
local term $-\Re\mu$ cancels the boundary contribution
$[\Re\mu]_{\rm in}^{\rm out}$ in $\Delta_1$; for an incident plane wave
$\Re\mu_{\rm in}=0$. We now carry out the analogous calculation for
linearized gravity.

Use tetrad indices $1=k$, $2=n$, $3=m$, $4=\bar m$, with
$T_{ab\ldots}$ denoting the contraction of a tensor with the tetrad
vectors after differentiation. For helicity $+2$,
\begin{equation}
h_{\mu\nu}=e^{iS/\epsilon}
\left(b_{0\mu\nu}+\epsilon b_{1\mu\nu}+\mathcal O(\epsilon^2)\right),
\qquad
b_{0\mu\nu}=\mathcal B\,m_\mu m_\nu .
\end{equation}
The only nonvanishing component of $b_0$ is $b_{0\,44}=\mathcal B$, and the
TT component of $b_1$ is $b_{1\,44}=i\mathcal B\varphi_2$. In vacuum the
linearized Weyl tensor equals the linearized Riemann tensor, so that,
up to background-curvature terms of relative order $\epsilon^2$,
\begin{equation}
\delta C_{n\bar m n\bar m}=\delta R_{2424}
=\frac12\left(
\nabla_2\nabla_4 h_{24}+\nabla_4\nabla_2 h_{24}
-\nabla_4\nabla_4 h_{22}-\nabla_2\nabla_2 h_{44}
\right).
\label{eq:deltaC_expansion}
\end{equation}

Each term follows from
\begin{equation}
\nabla_a\nabla_bh_{cd}=e^{iS/\epsilon}\Big[-\frac{1}{\epsilon^2}k_ak_bb_{cd}
+\frac{i}{\epsilon}\left(\nabla_ak_b\,b_{cd}+k_b\nabla_ab_{cd}+k_a\nabla_bb_{cd}\right)
+\mathcal O(1)\Big],
\end{equation}
with $b=b_0+\epsilon b_1$. Although $b_{0\,24}$ and $b_{0\,22}$ vanish,
their derivatives need not, because the frame varies. In the first two
terms of Eq.~\eqref{eq:deltaC_expansion} the $\epsilon^{-2}$ term vanishes
because $k\cdot\bar m=0$; in the $\epsilon^{-1}$ bracket only the
contraction $k\cdot n=-1$ survives, and
$\bar m^b\nabla_b b_{0cd}\,n^c\bar m^d=\mathcal B\,n\cdot\nabla_{\bar m}m=\mathcal B\bar\mu$.
Hence
\begin{equation}
\nabla_2\nabla_4h_{24}+\nabla_4\nabla_2h_{24}
=e^{iS/\epsilon}\left[-\frac{2i}{\epsilon}\,\mathcal B\bar\mu+\mathcal O(1)\right].
\end{equation}
The third term is $\mathcal O(1)$, since $b_{0\,22}=0$ and $\bar m\cdot k=0$.
In the last term,
$(n\cdot\nabla b_0)_{44}=n\cdot\nabla\mathcal B+2i\mathcal B A_n$, and
\begin{equation}
\nabla_2\nabla_2h_{44}
=e^{iS/\epsilon}\left[
-\frac{\mathcal B}{\epsilon^2}
-\frac{i}{\epsilon}\left(\mathcal B\varphi_2+2\,n\cdot\nabla\mathcal B
-\mathcal B\,n^an^b\nabla_ak_b+4i\mathcal B A_n\right)
+\mathcal O(1)\right].
\end{equation}
Collecting terms,
\begin{equation}
\delta C_{n\bar m n\bar m}
=\frac{\mathcal B}{2\epsilon^2}\,e^{iS/\epsilon}
\Big[1+i\epsilon\left(\varphi_2+2\,n\cdot\nabla\ln\mathcal B
-n^an^b\nabla_ak_b-2\Re\mu\right)
-\epsilon\left(4A_n+2\Im\mu\right)+\mathcal O(\epsilon^2)\Big].
\label{eq:deltaC_result}
\end{equation}

On a curved background the linearized Weyl tensor changes under a gauge
transformation by $\pounds_\xi C_{abcd}$. For transformations compatible
with the WKB form, $\xi=\epsilon\,e^{iS/\epsilon}\xi_0$, this change is of
relative order $\epsilon^2$, so the phase in Eq.~\eqref{eq:deltaC_result}
is gauge invariant at the order considered.

The structure is that of Eq.~\eqref{eq:F_nm}, with the weight $s=2$ in
front of $n\cdot\nabla\ln\mathcal B-\Re\mu$. The terms
$2\,n\cdot\nabla\ln\mathcal B$ and $n^an^b\nabla_ak_b$ are local: they are
evaluated at the detector, do not accumulate along the ray, and decay in
the far zone. The term $-2\Re\mu$ cancels the boundary contribution
$2[\Re\mu]_{\rm in}^{\rm out}$ in Eq.~\eqref{eq:Delta2_Delta1}. This
cancellation is exact. It does not rely on the decay of $\Re\mu$, which
is guaranteed only at first order in $M$: at second order, on the
outgoing branch before the caustic, the right-hand side of
Eq.~\eqref{eq:dmu} is dominated by the approximately constant product
$\sigma\lambda$, and $\Re\mu$ grows with the distance from the lens. For both fields, therefore, the accumulated phase of the invariant
field strength, measured relative to the scalar wave, is
\begin{equation}
\Delta_s^{\rm inv}=\epsilon\,s^2\left(\tfrac12I_C-I_A\right),
\qquad s=1,2,
\label{eq:Delta_inv}
\end{equation}
up to local terms at the detector, and $\Delta_2^{\rm inv}=4\Delta_1^{\rm inv}$.
The non-TT part of $b_1$ does not affect this result. The bulk
calculation of Sec.~\ref{sec:schwarzschild_ray} and the partial-wave
calculation of Sec.~\ref{sec:partial_waves} therefore evaluate the same
invariant quantity.

\section{Schwarzschild: bulk evaluation}
\label{sec:schwarzschild_ray}

From now on we use coordinates in units of $M$, with $\epsilon=1/(M\omega)$
and $\rho=b/M$, and consider the weak-deflection regime $\rho\gg1$. The
turning point $r_m$ satisfies $r_m=b-M+\mathcal O(M^2/b)$, so the leading
coefficients below are the same whether expressed in $b$ or in $r_m$.

\subsection{Weyl integral}

In the parallel frame fixed on the incident screen, the weak-deflection
evaluation of Appendix~\ref{app:weyl_integral} gives
\begin{equation}
I_C=-\frac{3\pi}{8\rho^3}+\mathcal O(\rho^{-4}).
\label{eq:IC_result}
\end{equation}
The straight-line contribution at order $\rho^{-2}$ is a total derivative
and cancels between past and future infinity. The first nonvanishing
term comes from the bending of the trajectory.

\subsection{Connection integral}

Commuting derivatives in the transport of Eq.~\eqref{eq:connection} gives
\begin{equation}
\frac{\dd A_a}{\dd\lambda}=-(\nabla_ak^b)A_b+i\,C_{m\bar mka}.
\label{eq:A_transport}
\end{equation}
Since $A_a=\mathcal O(M)$ and $\nabla_ak^b=\mathcal O(M)$, the first term
is of second order, and $|A_m|^2$ can be computed to $\mathcal O(M^2)$
along the unperturbed ray. Take $k=\partial_t+\partial_z$, the ray
through $(x,y,z)=(b,0,\lambda)$, and $m=(\hat x+i\hat y)/\sqrt2$. In the
static frame the vacuum Weyl tensor is determined by its electric part,
$E_{ij}=(M/r^3)(\delta_{ij}-3n_in_j)$, and
\begin{equation}
C_{m\bar mkm}=E_{ij}\hat z^im^j=-\frac{3Mb\lambda}{\sqrt2\,r^5}.
\end{equation}
Integrating Eq.~\eqref{eq:A_transport} from the incident screen, where
$A_a=0$,
\begin{equation}
A_m=\frac{iMb}{\sqrt2\,r^3},
\qquad
I_A=\frac{M^2b^2}{2}\int_{-\infty}^{\infty}\frac{\dd\lambda}{(b^2+\lambda^2)^3}
=\frac{3\pi}{16\rho^3}+\mathcal O(\rho^{-4}).
\label{eq:IA_result}
\end{equation}
The identities $\Cm=E_{zz}$, $\Cm=-2\Re C_{n\bar mkm}$ and
$C_{m\bar mkm}=E_{zm}$ used here have been checked numerically with the
full vacuum Weyl tensor.

\subsection{Result}

Combining Eqs.~\eqref{eq:h2_law}, \eqref{eq:IC_result} and
\eqref{eq:IA_result},
\begin{equation}
\Delta_s=\epsilon\,s^2\left(-\frac{3\pi}{16}-\frac{3\pi}{16}\right)\rho^{-3}
=-\frac{3\pi}{8}\,\frac{s^2}{M\omega\,\rho^3}+\mathcal O\!\left(\frac{1}{M\omega\,\rho^4}\right).
\label{eq:ray_result}
\end{equation}
The Weyl integral contributes, through the Ricci identity, one half of
the electromagnetic coefficient; the connection integral contributes the
other half. The gravitational coefficient is four times the
electromagnetic one.

\section{Schwarzschild: partial-wave evaluation}
\label{sec:partial_waves}

We now compute the same quantity from the scattering amplitudes. Nothing
in this section depends on a choice of frame, tetrad, or gauge.

\subsection{Constraint at first order in $M$}

In the static linearized field $h_{\mu\nu}=2U\,{\rm diag}(1,1,1,1)$, with
$U=M/r$, the Born amplitude is determined by
$h_{\mu\nu}T^{\mu\nu}=2U(T^{00}+T^{ii})$. For a massless scalar
$T^{00}+T^{ii}\to2\omega^2$, independent of the angle. For the Maxwell
field $T^{00}+T^{ii}=2T^{00}\to2\omega^2\cos^2(\theta/2)$ between states
of equal helicity. For gravitons the helicity-preserving amplitude
carries $\cos^4(\theta/2)$ \cite{DeLogiKovacs1977,Bautista2023}. Hence
\begin{equation}
f_s^{(1)}(\theta)=\cos^{2s}\!\tfrac\theta2\,f_0^{(1)}(\theta),
\qquad
\cos^2\tfrac\theta2=1-\frac{q^2}{4\omega^2},
\label{eq:Born}
\end{equation}
where $q=2\omega\sin(\theta/2)$. Since $f_0^{(1)}\propto\omega^2/q^2$,
the difference $f_s^{(1)}-f_0^{(1)}$ is a polynomial in $q^2$. It is a
contact term in impact-parameter space. At first order in $M$ there is
therefore no spin-dependent phase at $b\neq0$, to all orders in $1/\omega$.

\subsection{Phase shifts and angular functions}

In Schwarzschild spacetime the helicity-preserving amplitude takes the
standard form \cite{FuttermanHandlerMatzner1988,Dolan2008,Crispino2009}
\begin{equation}
f_s(\theta)=\frac{1}{2i\omega}\sum_l(2l+1)\,e^{2i\delta^{(s)}_l}\,d^l_{ss}(\theta)
\label{eq:partial_wave}
\end{equation}
The phase shifts
$\delta^{(s)}_l$ are those of the Regge--Wheeler equation
$\dd^2\psi/\dd r_*^2+(\omega^2-V_s)\psi=0$, with
\begin{equation}
V_s=f\left[\frac{l(l+1)}{r^2}+(1-s^2)\frac{2M}{r^3}\right],
\qquad f=1-\frac{2M}{r}.
\label{eq:RW}
\end{equation}
For $s=2$ the axial and polar phase shifts are related by Chandrasekhar's
transformation \cite{Chandrasekhar1983}. Their relative phase is of order
$M\omega/l^4$, far beyond the order considered here, so helicity reversal
does not enter.

We count orders in $\epsilon=1/(M\omega)$ at fixed $\rho$, with
$L=l+\tfrac12=\omega b$, and keep all terms of first order in $\epsilon$.
These are the terms $(M\omega)^n/L^{n+1}=\epsilon/\rho^{n+1}$.

\textit{Radial part.} Relative to $\omega^2$, the spin-dependent part of
$V_s$ is of order $\epsilon^2$. Its effect on the phase shift is
therefore given exactly, at order $\epsilon$, by first-order WKB
perturbation:
\begin{equation}
2\delta^{(s)}_l-2\delta^{(0)}_l
=2s^2M\int_{r_t}^{\infty}\frac{\dd r}{r^3\sqrt{\omega^2-fL^2/r^2}}
=\frac{2s^2M}{\omega b^2}\,\mathcal J\!\left(\frac Mb\right),
\label{eq:radial}
\end{equation}
with
\begin{equation}
\mathcal J(\mu)=\int_0^{u_t}\frac{u\,\dd u}{\sqrt{1-u^2+2\mu u^3}}
=1+\frac{3\pi}{4}\,\mu+\mathcal O(\mu^2),
\label{eq:J}
\end{equation}
where $u_t$ is the root near unity. We obtained the coefficient $3\pi/4$
numerically, to ten significant figures, by Richardson extrapolation.
Higher-order WKB corrections and the Langer term are spin independent at
this order.

\textit{Angular part.} The functions $d^l_{ss}$ obey an equation whose
potential differs from that of $P_l$ by $s^2/\cos^2(\theta/2)$
(Appendix~\ref{app:angular}). In the oscillatory region $L\theta\gg1$,
their phase differs from that of $P_l$ by
\begin{equation}
\Phi^{\rm ang}_s-\Phi^{\rm ang}_0=-\frac{s^2}{L}\tan\frac\theta2 .
\label{eq:angular}
\end{equation}

\subsection{Stationary phase}

For attractive scattering the dominant branch of Eq.~\eqref{eq:partial_wave}
is $e^{2i\delta_l+iL\theta}$, and the stationary point is fixed by
$\theta=-2\,\dd\delta_l/\dd L$. At fixed scattering angle, and to first
order in the spin-dependent pieces, the displacement of the stationary
point does not affect the phase. Therefore
\begin{equation}
\arg f_s(\theta)-\arg f_0(\theta)
=\frac{s^2}{\omega}\left[\frac{2M}{b^2}\,\mathcal J\!\left(\frac Mb\right)
-\frac1b\tan\frac{\theta(b)}{2}\right],
\label{eq:stationary}
\end{equation}
with the null deflection angle
\begin{equation}
\theta(b)=\frac{4M}{b}+\frac{15\pi}{4}\frac{M^2}{b^2}+\mathcal O(M^3/b^3).
\end{equation}
The terms of first order in $M$ cancel, $2M/b^2-2M/b^2=0$, as required by
Eq.~\eqref{eq:Born}. At second order,
\begin{equation}
\arg f_s-\arg f_0
=s^2\left(\frac{3\pi}{2}-\frac{15\pi}{8}\right)\frac{1}{M\omega\,\rho^3}
=-\frac{3\pi}{8}\,\frac{s^2}{M\omega\,\rho^3}
+\mathcal O\!\left(\frac{1}{M\omega\,\rho^4},\epsilon^3\right),
\label{eq:pw_result}
\end{equation}
in exact agreement with Eq.~\eqref{eq:ray_result}. The factor $s^2$
enters here through the coefficient $(1-s^2)$ of the Regge--Wheeler
potential and through the angular potential $s^2/\cos^2(\theta/2)$.

Equation~\eqref{eq:stationary} holds at first order in $\epsilon$ for
all $\rho$. Its values, computed with the exact deflection angle and the
exact integral $\mathcal J$, are listed in Table~\ref{tab:invariant}.
They approach $-3\pi/8$ with a correction of order $\rho^{-1}$; a
Richardson extrapolation of the last two entries gives $-1.178088$.

\begin{table}[t]
\caption{Invariant post-eikonal phase difference in Schwarzschild,
Eq.~\eqref{eq:stationary}, evaluated with the exact deflection angle
$\theta(b)$ and the exact integral $\mathcal J$. The last column is
$M\omega\,\rho^3(\arg f_s-\arg f_0)/s^2$, whose limit is
$-3\pi/8=-1.178097\ldots$}
\label{tab:invariant}
\begin{ruledtabular}
\begin{tabular}{rrrr}
$\rho=b/M$ & $\mathcal J(1/\rho)$ & $\theta(b)$ & $M\omega\rho^3\Delta_s/s^2$\\
\hline
10   & 1.369419408 & 0.5903957876 & $-3.019852$ \\
20   & 1.142697847 & 0.2361359954 & $-1.739966$ \\
40   & 1.064444142 & 0.1081040765 & $-1.412052$ \\
80   & 1.030766135 & 0.0519284942 & $-1.285951$ \\
160  & 1.015046466 & 0.0254708775 & $-1.229990$ \\
400  & 1.005940974 & 0.0100743045 & $-1.198394$ \\
800  & 1.002957804 & 0.0050184915 & $-1.188171$ \\
1600 & 1.001475754 & 0.0025046124 & $-1.183116$ \\
3200 & 1.000737093 & 0.0012511518 & $-1.180602$ \\
\end{tabular}
\end{ruledtabular}
\end{table}

\subsection{Numerical checks}

We integrated the Regge--Wheeler equations for $s=0$ and $s=1$
numerically. Each integration starts inside the centrifugal barrier with
the growing WKB solution and continues to $r=4000M$, where the phase is
read off with the local wave number. The resulting phase-shift
differences are compared with Eq.~\eqref{eq:radial} in
Table~\ref{tab:ODE}. The relative differences are between
$4\times10^{-6}$ and $5\times10^{-5}$. At $b\simeq20M$ the second-order
part of Eq.~\eqref{eq:radial} accounts for about $12\%$ of the total,
so it is confirmed to better than $10^{-3}$. The angular phase
\eqref{eq:angular} was checked against the zeros of $d^l_{11}$ and
$d^l_{22}$ relative to those of $P_l$, for $l=400$ and
$\theta=0.05,\,0.2$, with agreement at the level of $10^{-4}$.
The Python scripts used for these checks and for Table~\ref{tab:invariant}
were written with the assistance of an AI tool (see the
Acknowledgments). They use standard libraries (SciPy and mpmath) and are
provided upon request.

\begin{table}[t]
\caption{Phase-shift difference $2(\delta^{(1)}_l-\delta^{(0)}_l)$ from
direct integration of the Regge--Wheeler equations, compared with the
first-order WKB expression, Eq.~\eqref{eq:radial}.}
\label{tab:ODE}
\begin{ruledtabular}
\begin{tabular}{rrrrrr}
$M\omega$ & $l$ & $b/M$ & numerical & Eq.~\eqref{eq:radial} & relative difference\\
\hline
10 & 200 & 20.05 & $5.682852\times10^{-4}$ & $5.682874\times10^{-4}$ & $-3.8\times10^{-6}$\\
10 & 400 & 40.05 & $1.327065\times10^{-4}$ & $1.327125\times10^{-4}$ & $-4.5\times10^{-5}$\\
20 & 400 & 20.02 & $2.849049\times10^{-4}$ & $2.849075\times10^{-4}$ & $-9.2\times10^{-6}$\\
\end{tabular}
\end{ruledtabular}
\end{table}

\section{Discussion}
\label{sec:discussion}

\subsection{Helicity-odd and helicity-even sectors}

The first post-eikonal correction separates cleanly under helicity
reversal. The helicity-odd part, carried by $G_s$, is linear in the
polarization connection and in $h$. It underlies the spin-optical
corrections to ray dynamics, including the spin Hall effect
\cite{Oancea2020,Andersson2021,FrolovShoom2024,FrolovKoek2025}. The
helicity-even part, measured relative to the scalar wave, is quadratic in
$h$ for both the electromagnetic and the gravitational field,
Eq.~\eqref{eq:h2_law}. The direct Weyl coupling does not produce a
separate gravitational structure. Through the Ricci identity it is
precisely what upgrades the optical term from $s$ to $s^2$. A description
of the gravitational wave obtained by weighting the electromagnetic
helicity-even sector with $h^2$ is therefore complete at this order,
whereas a weighting linear in $s$ would miss exactly the contribution of
the Weyl term.

The coefficient in Eq.~\eqref{eq:main_result} has a simple origin in
each calculation. Along the ray, half of it comes from the Weyl
integral and half from the transverse twist of the parallel frame. In
partial waves it is the difference between the second-order radial
term, $3\pi/2$, and the kinematic term produced by the angular functions
at the second-order deflection, $15\pi/8$.

\subsection{Relation to partial-wave and wave-optics studies}

Partial-wave treatments of scalar, electromagnetic and gravitational
scattering by black holes are well developed
\cite{FuttermanHandlerMatzner1988,Dolan2008,Crispino2009}. Wave-optics
studies of gravitational lensing by rotating black holes have recently
been carried out with the Teukolsky equation. In particular, Kubota
\textit{et al.}~\cite{Kubota2024} solved the Teukolsky equation for
gravitational waves emitted by a binary and lensed by a Kerr black hole,
in the regime where the wavelength is comparable to the gravitational
radius. They found a helicity-dependent, small-period oscillation in the
power spectrum of the forward amplification factor, which grows with the
black-hole spin for prograde configurations. That work addresses the
full wave problem numerically, in a regime where $M\omega$ is of order
unity, and its central effect distinguishes the two helicities. The
present work addresses a different question. We work in the
high-frequency, weak-deflection regime, isolate the helicity-even
correction at first post-eikonal order, and compare fields of different
spin on the same background. Our result is an analytic coefficient that
depends on $s^2$ and is the same for both helicities. Low-frequency
comparisons between Teukolsky scattering and scattering amplitudes
\cite{Bautista2023} are also complementary to ours.

Two further lines of work deserve comment. Beyond-geometric-optics
corrections to gravitational lensing have been studied for gravitational
waves at linear order in the lens potential \cite{CusinLagos2020} and,
for scalar waves, up to second order in $G$ \cite{BruyerePitrou2026}. In
the scalar case the first correction appears at order $G^2$ and grows
with the distance from the lens. This is the behavior of the common term
$\Box\mathcal B/(2\mathcal B)$, which cancels in $\Delta_s$. The
spin-dependent integrands considered here decay as $r^{-6}$, and their
integrals converge. The absence of a spin-dependent phase at order $G$ is
consistent with the linear-order analysis of Ref.~\cite{CusinLagos2020}.

In the scattering-amplitude literature, the bending angle of massless
particles has been computed at one loop, and it contains spin-dependent
terms of the form $bu^S\,\hbar G^2M/b^3$
\cite{BjerrumBohr2015,BjerrumBohr2016}. These terms originate in quantum
loops, are linear in $M$, and are obtained for $\hbar\omega\ll M$ at
first order in $\hbar$. The effect found here is of a different nature.
It is a classical wave effect, proportional to $G^2M^2/\omega$. In
particle variables, with $E=\hbar\omega$, it contributes to the
deflection angle only at order $\hbar^2$, beyond the truncation of
Refs.~\cite{BjerrumBohr2015,BjerrumBohr2016}. At the classical level the
helicity-preserving eikonal phase of general relativity is universal at
second post-Minkowskian order \cite{AccettulliHuber2020}. Finally, a
helicity-even response of gravitons has recently been identified in a
quantum-geometric gradient expansion \cite{Mehraeen2026}. That effect is
sourced by the Ricci tensor and vanishes in vacuum, whereas the effect
described here arises entirely from the Weyl curvature.

\subsection{Magnitude of the effect}

For a gravitational wave, Eq.~\eqref{eq:main_result} gives
$|\Delta_2|=3\pi/(2M\omega\rho^3)$. With $GM_\odot/c^3\simeq4.9\times10^{-6}$~s,
a wave of frequency $f=100$~Hz passing a black hole of $10^6M_\odot$ has
$M\omega\simeq3\times10^3$. At $\rho=10$ this gives
$|\Delta_2|\sim10^{-6}$~rad, and at $\rho=100$, $|\Delta_2|\sim10^{-9}$~rad. For a
black hole of $10^8M_\odot$ the corresponding values are $10^{-8}$ and
$10^{-11}$~rad. At $\rho=10$ the higher-order terms in $1/\rho$ increase these
numbers by a factor of about $2.6$ (Table~\ref{tab:invariant}), which does
not change the conclusion. For cosmological lenses the relevant impact
parameters are of the order of the Einstein radius, $b/M\gtrsim10^9$, and
the effect is entirely negligible. Larger values would require
$M\omega\lesssim1$, where the post-eikonal expansion no longer applies and
the full wave problem must be solved \cite{Kubota2024}. The helicity-even
phase difference is therefore far below the phase resolution of present
and planned detectors. It also enters the observed frequency-dependent
phase together with the spin-independent term
$\Box\mathcal B/(2\mathcal B)$, which is larger and grows with the distance
from the lens \cite{BruyerePitrou2026}.

The interest of the result is thus structural rather than observational.
Equation~\eqref{eq:h2_law} fixes how the helicity-even part of the first
post-eikonal correction depends on the spin of the field, and it
complements the helicity-odd spin-optical sector. The coefficient
$-3\pi s^2/8$, together with the vanishing of the first-order term, also
provides an analytic benchmark for numerical treatments of wave
propagation in black-hole spacetimes, such as Teukolsky or
wave-optics lensing codes, in the high-frequency regime.

\subsection{Kerr}

The derivation of Sec.~\ref{sec:law} uses only the vacuum Einstein
equations and a parallel tetrad, so Eq.~\eqref{eq:h2_law} holds in Kerr
spacetime as well. In the weak-deflection expansion the rotation
parameter first enters the Weyl integral at order $\rho^{-4}$
(Appendix~\ref{app:weyl_integral}). Gravitomagnetic corrections to $A_m$
are of relative order $a/b$, so they also affect $I_A$ only at order
$\rho^{-4}$. The leading coefficient $-3\pi s^2/8$ is therefore
expected to be unchanged for $a\neq0$. An invariant confirmation, and the
rotation-dependent terms at order $\rho^{-4}$, require an invariant
evaluation, for example from the
Teukolsky equation with spin-weighted spheroidal harmonics. This is left
for future work.

\section{Conclusions}
\label{sec:conclusions}

We have studied the helicity-even part of the first post-eikonal
correction for scalar, electromagnetic, and linearized gravitational
waves on Ricci-flat backgrounds. The direct Weyl coupling acts on the
circular TT states as $2\Cm$ times the identity, so it neither converts
nor splits the helicities. In a parallel null tetrad the Ricci identity
for $\mu$ rewrites $\Cm$ as a total derivative plus the optical
combination $-2\Re(\rho\mu+\sigma\lambda)$. With this identity the
helicity-even phase of a spin-$s$ wave, measured relative to the scalar
wave, equals $s^2$ times a common bulk integral plus endpoint terms; for
the invariant field strengths the endpoint terms cancel, apart from
local terms at the detector. In the gravitational case the
Weyl coupling supplies the part of the $s^2$ scaling that the
frame-derivative terms alone would miss.

For weak-deflection scattering by a Schwarzschild black hole the
helicity-even phase difference vanishes at first order in $M$ and at
second order is $\Delta_s=-3\pi s^2/(8M\omega\rho^3)$. We obtained this
coefficient in two ways: by integrating the bulk quantities along the
ray, and from the Regge--Wheeler phase shifts together with the
asymptotics of $d^l_{ss}$. The second route involves no frame or gauge
choice, and direct numerical integration of the radial equations
confirms it. The spin-optical and post-eikonal corrections of
electromagnetic and gravitational waves are therefore organized by the
first and second powers of the helicity. Extending the invariant
evaluation to the rotation-dependent terms in Kerr spacetime remains
open.

\begin{acknowledgments}
The author wishes to express sincere gratitude to the Conselho Nacional de Desenvolvimento Cient\'{i}fico e Tecnol\'{o}gico (CNPq) and Funda\c{c}\~{a}o Cearense de Apoio ao Desenvolvimento Cient\'{i}fico e Tecnol\'{o}gico (FUNCAP) for their valuable support through grants Project UNI-00210-00230.01.00/23 (FUNCAP), 303438/2026-6 (CNPq/PQ) and  420854/2025-8 (CNPq).

\textit{Use of AI tools.}---This work made use of Claude (Anthropic, model claude-opus-5-5), accessed via the Claude desktop application in September 2026. Based on author prompts, the tool was used to assist with analytical derivations, numerical checks via Python scripts, and drafting and editing the manuscript. All analytical results were independently verified by the author, and all numerical outputs were reproduced by rerunning the scripts. The author takes full responsibility for the content of this paper.

\end{acknowledgments}
\section*{Data Availability Statement}
The code used in this work is available from the corresponding author upon reasonable request.
\section*{Conflict of Interest Statement}
The author declares that there is no conflict of interest regarding the publication of this paper.

\appendix

\section{Weak-deflection evaluation of the Weyl integral}
\label{app:weyl_integral}

We work in units of $M$ and normalize the energy of the reference null
geodesic to unity. For Schwarzschild, and for the equatorial Kerr
problem with $\alpha=a/M$, the radial equation reads
$(\dd\rho/\dd\lambda)^2=B(\rho)$, with
\begin{equation}
B(\rho)=1-\frac{\ell^2-\alpha^2}{\rho^2}+\frac{2(\ell-\alpha)^2}{\rho^3}.
\end{equation}
Here $\ell$ is the impact parameter in units of $M$, and the Schwarzschild
case is $\alpha=0$. The turning point $\rho_m$, defined by $B(\rho_m)=0$,
satisfies $\ell=\rho_m+1+\mathcal O(\rho_m^{-1})$. With $x=\rho/\rho_m$,
\begin{equation}
B=B_0+\rho_m^{-1}B_1+\mathcal O(\rho_m^{-2}),
\qquad
B_0=1-\frac{1}{x^2},
\qquad
B_1=2\left(\frac{1}{x^3}-\frac{1}{x^2}\right),
\end{equation}
so $\alpha$ does not enter at this order. The affine distance from the
turning point is
\begin{equation}
\Lambda(\rho)=\int_{\rho_m}^\rho\frac{\dd\rho'}{\sqrt{B}}
=\rho_mL_0(x)+L_1(x)+\mathcal O(\rho_m^{-1}),
\qquad
L_0=\sqrt{x^2-1},\quad L_1=\sqrt{\frac{x-1}{x+1}} .
\end{equation}

In the parallel frame fixed on the incident screen, the quantity
$\kappa=(\rho^2+\alpha^2)e_2^\phi-\alpha e_2^t$ obeys
$\dd\kappa/\dd\lambda=-1$. Denote by $\Lambda_m$ its value at the turning
point. Then $\kappa=\Lambda_m\pm\Lambda(\rho)$ on the incoming and outgoing
branches, and
\begin{equation}
\Cm=\frac{\rho^2-3[\Lambda_m\pm\Lambda(\rho)]^2}{\rho^5}.
\end{equation}
Adding both branches at the same radius cancels the terms linear in
$\Lambda\Lambda_m$:
\begin{equation}
I_C=2\int_{\rho_m}^\infty
\frac{\rho^2-3(\Lambda^2+\Lambda_m^2)}{\rho^5\sqrt{B}}\,\dd\rho .
\end{equation}
Since $\Lambda_m=\mathcal O(1)$, its contribution enters at
$\mathcal O(\rho_m^{-4})$. More precisely, a parallel null rotation
$e_2\to e_2+[\delta/(\ell-\alpha)]\,k$ changes the integral by exactly
$-3(2\Lambda_m\delta+\delta^2)J_5$, with
$J_5=2\int_{\rho_m}^\infty\dd\rho/(\rho^5\sqrt B)=4/(3\rho_m^4)+\dots$

Expanding, $I_C=I_0\rho_m^{-2}+I_1\rho_m^{-3}+\mathcal O(\rho_m^{-4})$. The
leading coefficient,
\begin{equation}
I_0=2\int_1^\infty\frac{3-2x^2}{x^4\sqrt{x^2-1}}\,\dd x=0,
\end{equation}
vanishes because, along the straight line $\rho^2=\rho_m^2+\lambda^2$,
\begin{equation}
\Cm\simeq\frac{\rho_m^2-2\lambda^2}{(\rho_m^2+\lambda^2)^{5/2}}
=\frac{\dd}{\dd\lambda}\left[\frac{\lambda}{(\rho_m^2+\lambda^2)^{3/2}}\right].
\end{equation}
This is the explicit form of the boundary term $-2[\Re\mu]$ in
Eq.~\eqref{eq:Ricci}. At the next order,
\begin{equation}
I_1=-2\int_1^\infty
\frac{6(x-1)+\dfrac{2x^2-3}{x(x+1)}}{x^4\sqrt{x^2-1}}\,\dd x
=-\frac{3\pi}{8},
\end{equation}
which gives Eq.~\eqref{eq:IC_result}.

A numerical evaluation of the same integral for the equatorial Kerr
problem with $a/M=0.5$, with $\ell$ determined from the turning radius
and both branches integrated to infinity, gives
$\rho_m^3I_C=-2.071951$, $-1.554767$, $-1.352459$, $-1.262133$,
$-1.219368$, $-1.194431$, $-1.186235$ and $-1.182159$ for
$\rho_m=10$, $20$, $40$, $80$, $160$, $400$, $800$ and $1600$. These values converge
to $-3\pi/8$, consistently with the absence of $\alpha$ at this order.

\section{Angular phase of $d^l_{ss}$}
\label{app:angular}

The rotation functions with equal indices can be written in terms of
Jacobi polynomials,
\begin{equation}
d^l_{ss}(\theta)=\cos^{2s}\!\tfrac\theta2\;P^{(0,2s)}_{l-s}(\cos\theta),
\end{equation}
and the Hilb-type asymptotics of Jacobi polynomials \cite{Szego1939}
give $d^l_{ss}(\theta)\simeq\sqrt{\theta/\sin\theta}\,J_0(L\theta)$ at
leading order, with $L=l+\tfrac12$ for every $s$. The spin dependence
appears at the next order. The function $\sqrt{\sin\theta}\,d^l_{ss}$ obeys
\begin{equation}
\frac{\dd^2u}{\dd\theta^2}
+\left[L^2+\frac{1}{4\sin^2\theta}-\frac{s^2}{\cos^2(\theta/2)}\right]u=0,
\end{equation}
which reduces to the Legendre case for $s=0$. In the oscillatory region
the WKB phase difference is
\begin{equation}
\Phi^{\rm ang}_s-\Phi^{\rm ang}_0
\simeq-\frac{s^2}{2L}\int_0^\theta\frac{\dd\theta'}{\cos^2(\theta'/2)}
=-\frac{s^2}{L}\tan\frac\theta2 .
\end{equation}
The integration constant is fixed by the small-angle expansion
$d^l_{ss}=1-[l(l+1)-s^2]\theta^2/4+\dots$, which corresponds to
$J_0(\sqrt{L^2-s^2}\,\theta)$. Corrections are of relative order $L^{-2}$
and $(L\theta)^{-2}$. In the scattering problem these correspond to
order $\epsilon^3$.

\bibliography{helicity_squared_references}

@article{Berry1984,
  author  = {Michael V. Berry},
  title   = {Quantal phase factors accompanying adiabatic changes},
  journal = {Proceedings of the Royal Society of London A},
  volume  = {392},
  pages   = {45--57},
  year    = {1984},
  doi     = {10.1098/rspa.1984.0023}
}

@article{Isaacson1968I,
  author  = {Richard A. Isaacson},
  title   = {Gravitational Radiation in the Limit of High Frequency. I. The Linear Approximation and Geometrical Optics},
  journal = {Physical Review},
  volume  = {166},
  pages   = {1263--1271},
  year    = {1968},
  doi     = {10.1103/PhysRev.166.1263}
}

@article{Isaacson1968II,
  author  = {Richard A. Isaacson},
  title   = {Gravitational Radiation in the Limit of High Frequency. II. Nonlinear Terms and the Effective Stress Tensor},
  journal = {Physical Review},
  volume  = {166},
  pages   = {1272--1280},
  year    = {1968},
  doi     = {10.1103/PhysRev.166.1272}
}

@article{FrolovShoom2011,
  author  = {Valeri P. Frolov and Andrey A. Shoom},
  title   = {Spinoptics in a stationary spacetime},
  journal = {Physical Review D},
  volume  = {84},
  pages   = {044026},
  year    = {2011},
  doi     = {10.1103/PhysRevD.84.044026}
}

@article{Dolan2018,
  author  = {Sam R. Dolan},
  title   = {Geometrical optics for scalar, electromagnetic and gravitational waves on curved spacetime},
  journal = {International Journal of Modern Physics D},
  volume  = {27},
  pages   = {1843010},
  year    = {2018},
  doi     = {10.1142/S0218271818430101}
}

@misc{Dolan2018HO,
  author        = {Sam R. Dolan},
  title         = {Higher-order geometrical optics for electromagnetic waves on a curved spacetime},
  year          = {2018},
  eprint        = {1801.02273},
  archivePrefix = {arXiv},
  primaryClass  = {gr-qc}
}

@article{Yamamoto2018,
  author  = {Naoki Yamamoto},
  title   = {Spin Hall effect of gravitational waves},
  journal = {Physical Review D},
  volume  = {98},
  pages   = {061701},
  year    = {2018},
  doi     = {10.1103/PhysRevD.98.061701}
}

@article{Harte2019,
  author  = {Abraham I. Harte},
  title   = {Gravitational lensing beyond geometric optics: I. Formalism and observables},
  journal = {General Relativity and Gravitation},
  volume  = {51},
  pages   = {14},
  year    = {2019},
  doi     = {10.1007/s10714-018-2494-x}
}

@article{Oancea2020,
  author  = {Marius A. Oancea and J{\'e}r{\'e}mie Joudioux and I. Y. Dodin and D. E. Ruiz and Claudio F. Paganini and Lars Andersson},
  title   = {Gravitational spin Hall effect of light},
  journal = {Physical Review D},
  volume  = {102},
  pages   = {024075},
  year    = {2020},
  doi     = {10.1103/PhysRevD.102.024075}
}

@article{Frolov2020,
  author  = {Valeri P. Frolov},
  title   = {Maxwell equations in a curved spacetime: Spin optics approximation},
  journal = {Physical Review D},
  volume  = {102},
  pages   = {084013},
  year    = {2020},
  doi     = {10.1103/PhysRevD.102.084013}
}

@article{Andersson2021,
  author  = {Lars Andersson and J{\'e}r{\'e}mie Joudioux and Marius A. Oancea and Ayush Raj},
  title   = {Propagation of polarized gravitational waves},
  journal = {Physical Review D},
  volume  = {103},
  pages   = {044053},
  year    = {2021},
  doi     = {10.1103/PhysRevD.103.044053}
}

@article{Dahal2022,
  author  = {Pravin Kumar Dahal},
  title   = {Spin Optics for Gravitational Waves},
  journal = {Astronomy},
  volume  = {1},
  number  = {3},
  pages   = {271--287},
  year    = {2022},
  doi     = {10.3390/astronomy1030016}
}

@article{HarteOancea2022,
  author  = {Abraham I. Harte and Marius A. Oancea},
  title   = {Spin Hall effects and the localization of massless spinning particles},
  journal = {Physical Review D},
  volume  = {105},
  pages   = {104061},
  year    = {2022},
  doi     = {10.1103/PhysRevD.105.104061}
}

@article{AnderssonOancea2023,
  author  = {Lars Andersson and Marius A. Oancea},
  title   = {Spin Hall effects in the sky},
  journal = {Classical and Quantum Gravity},
  volume  = {40},
  number  = {15},
  pages   = {154002},
  year    = {2023},
  doi     = {10.1088/1361-6382/ace021}
}

@article{FrolovShoom2024,
  author  = {Valeri P. Frolov and Andrey A. Shoom},
  title   = {Gravitational spinoptics in a curved space-time},
  journal = {Journal of Cosmology and Astroparticle Physics},
  volume  = {2024},
  number  = {10},
  pages   = {039},
  year    = {2024},
  doi     = {10.1088/1475-7516/2024/10/039}
}

@article{FrolovSchwarzschild2024,
  author  = {Valeri P. Frolov},
  title   = {Spinoptics in the Schwarzschild spacetime},
  journal = {Journal of Cosmology and Astroparticle Physics},
  volume  = {2024},
  number  = {12},
  pages   = {051},
  year    = {2024},
  doi     = {10.1088/1475-7516/2024/12/051}
}

@article{FrolovKoek2025,
  author  = {Valeri P. Frolov and Alex Koek},
  title   = {Spinoptics in the Kerr spacetime: Polarized wave scattering},
  journal = {Physical Review D},
  volume  = {111},
  pages   = {104081},
  year    = {2025},
  doi     = {10.1103/PhysRevD.111.104081}
}

@misc{Ito2026,
  author        = {Ritsuki Ito and Kazuya Mameda and Naoki Yamamoto},
  title         = {Spin Hall effect and Berry curvature of gravitons from quantum field theory},
  year          = {2026},
  eprint        = {2605.19817},
  archivePrefix = {arXiv},
  primaryClass  = {gr-qc},
  note          = {Preprint}
}

@article{NewmanPenrose1962,
  author  = {Ezra T. Newman and Roger Penrose},
  title   = {An Approach to Gravitational Radiation by a Method of Spin Coefficients},
  journal = {Journal of Mathematical Physics},
  volume  = {3},
  pages   = {566--578},
  year    = {1962},
  doi     = {10.1063/1.1724257}
}

@book{Chandrasekhar1983,
  author    = {Subrahmanyan Chandrasekhar},
  title     = {The Mathematical Theory of Black Holes},
  publisher = {Clarendon Press},
  address   = {Oxford},
  year      = {1983},
  isbn      = {0198512910}
}

@article{DeLogiKovacs1977,
  author  = {William K. De Logi and Sandor J. Kov{\'a}cs},
  title   = {Gravitational scattering of zero-rest-mass plane waves},
  journal = {Physical Review D},
  volume  = {16},
  pages   = {237--244},
  year    = {1977},
  doi     = {10.1103/PhysRevD.16.237}
}

@book{FuttermanHandlerMatzner1988,
  author    = {J. A. H. Futterman and F. A. Handler and R. A. Matzner},
  title     = {Scattering from Black Holes},
  publisher = {Cambridge University Press},
  address   = {Cambridge},
  year      = {1988}
}

@article{Dolan2008,
  author  = {Sam R. Dolan},
  title   = {Scattering and absorption of gravitational plane waves by rotating black holes},
  journal = {Classical and Quantum Gravity},
  volume  = {25},
  pages   = {235002},
  year    = {2008},
  doi     = {10.1088/0264-9381/25/23/235002}
}

@article{Crispino2009,
  author  = {Lu{\'i}s C. B. Crispino and Sam R. Dolan and Ednilton S. Oliveira},
  title   = {Electromagnetic Wave Scattering by {Schwarzschild} Black Holes},
  journal = {Physical Review Letters},
  volume  = {102},
  pages   = {231103},
  year    = {2009},
  doi     = {10.1103/PhysRevLett.102.231103}
}

@article{Kubota2024,
  author  = {Kei-ichiro Kubota and Shun Arai and Hayato Motohashi and Shinji Mukohyama},
  title   = {Spin wave optics for gravitational waves lensed by a {Kerr} black hole},
  journal = {Physical Review D},
  volume  = {110},
  pages   = {124011},
  year    = {2024},
  doi     = {10.1103/PhysRevD.110.124011}
}

@article{Bautista2023,
  author  = {Yilber F. Bautista and Alfredo Guevara and Chris Kavanagh and Justin Vines},
  title   = {Scattering in black hole backgrounds and higher-spin amplitudes. {Part I}},
  journal = {Journal of High Energy Physics},
  volume  = {2023},
  number  = {03},
  pages   = {136},
  year    = {2023},
  doi     = {10.1007/JHEP03(2023)136}
}

@book{Szego1939,
  author    = {G{\'a}bor Szeg{\H o}},
  title     = {Orthogonal Polynomials},
  series    = {American Mathematical Society Colloquium Publications},
  volume    = {23},
  publisher = {American Mathematical Society},
  address   = {Providence, RI},
  year      = {1939}
}

@article{CusinLagos2020,
  author  = {Giulia Cusin and Macarena Lagos},
  title   = {Gravitational wave propagation beyond geometric optics},
  journal = {Physical Review D},
  volume  = {101},
  pages   = {044041},
  year    = {2020},
  doi     = {10.1103/PhysRevD.101.044041}
}

@misc{BruyerePitrou2026,
  author        = {Emma Bruy{\`e}re and Cyril Pitrou},
  title         = {Gravitational lensing beyond the eikonal approximation},
  year          = {2026},
  eprint        = {2601.10239},
  archivePrefix = {arXiv},
  primaryClass  = {gr-qc}
}

@article{BjerrumBohr2015,
  author  = {N. E. J. Bjerrum-Bohr and John F. Donoghue and Barry R. Holstein and Ludovic Plant{\'e} and Pierre Vanhove},
  title   = {Bending of Light in Quantum Gravity},
  journal = {Physical Review Letters},
  volume  = {114},
  pages   = {061301},
  year    = {2015},
  doi     = {10.1103/PhysRevLett.114.061301}
}

@article{BjerrumBohr2016,
  author  = {N. E. J. Bjerrum-Bohr and John F. Donoghue and Barry R. Holstein and Ludovic Plant{\'e} and Pierre Vanhove},
  title   = {Light-like scattering in quantum gravity},
  journal = {Journal of High Energy Physics},
  volume  = {2016},
  number  = {11},
  pages   = {117},
  year    = {2016},
  doi     = {10.1007/JHEP11(2016)117}
}

@article{AccettulliHuber2020,
  author  = {Manuel Accettulli Huber and Andreas Brandhuber and Stefano De Angelis and Gabriele Travaglini},
  title   = {Eikonal phase matrix, deflection angle, and time delay in effective field theories of gravity},
  journal = {Physical Review D},
  volume  = {102},
  pages   = {046014},
  year    = {2020},
  doi     = {10.1103/PhysRevD.102.046014}
}

@misc{Mehraeen2026,
  author        = {M. Mehraeen},
  title         = {Quantum geometry of gravitons},
  year          = {2026},
  eprint        = {2609.15237},
  archivePrefix = {arXiv}
}

@misc{Dahal2023,
  author        = {Pravin Kumar Dahal},
  title         = {Light propagation in {Kerr} spacetime},
  year          = {2023},
  eprint        = {2301.08250},
  archivePrefix = {arXiv},
  primaryClass  = {gr-qc}
}

\end{document}